\documentstyle[aps,prl,multicol,epsfig]{revtex}

\title{RNA secondary structure formation: a solvable model of 
heteropolymer folding}

\author{R.~Bundschuh~and T.~Hwa}
\address{Department of Physics,
University of California at San Diego,
La Jolla, CA  92093-0319, U.S.A.}

\date{March 4, 1999}

\draft

\begin{document}

\maketitle

\begin{abstract}
The statistical mechanics of heteropolymer structure formation is
studied in the context of RNA secondary structures. A designed RNA
sequence biased energetically towards a particular native structure (a
hairpin) is used to study the transition between the native and molten
phase of the RNA as a function of temperature. The transition is
driven by a competition between the energy gained from the polymer's
overlap with the native structure and the entropic gain of forming
random contacts.  A simplified G\=o-like model is proposed and solved
exactly. The predicted critical behavior is verified via exact
numerical enumeration of a large ensemble of similarly designed
sequences.
\end{abstract}

\begin{multicols}{2}
\narrowtext

A biopolymer such as a DNA or protein is a heteropolymer. It consists
of different types of monomers connected linearly in a specific order.
Interactions among the monomers give each polymer a robust three
dimensional structure on which its biological function depends.  This
sequence-to-structure relation is rather simple in the case of
complementary DNA strands, but can be very complex in the case of
proteins.  The latter has been intensively studied in the last decade
using many different approaches~\cite{dill95,woly97,gare97,shak97}.

A number of important ingredients are involved in determining the
structure of a heteropolymer. They include (i) thermal fluctuations
which ``denature'' the polymer into a random coil at high
temperatures, (ii) monomer-specific binding which freezes a random
heteropolymer into a ``glass'' at low temperatures, and (iii)~sequence
correlation which biases the polymer into a certain specific
(nonrandom) structure, commonly referred to as the ``native''
structure. The native structures are selected in nature by evolution,
but can also be obtained artificially through {\em sequence
design}~\cite{pand94,deut96}.  The interplay of these ingredients
leads to a number of phases depending on the environment (e.g., the
temperature) and the extent of sequence correlations or design.  The
nature of these phases and the transitions among them have been
discussed in the context of protein folding~\cite{woly97,gare97}.
However, protein-like models are {\em not} ideal systems to study
phase-related issues because proteins are rather short (typically
under $500$ monomers), and their thermodynamic limit is ambiguous:
simply taking longer proteins leads to problems that are neither
interesting nor relevant.

Here, we shall take the view that the statistical mechanics of a long
heteropolymer governed by the competing interactions mentioned above
are of interest in their own right, regardless of their immediate
application to protein folding.  We will study the molecule RNA, an
interesting biopolymer which has a mixture of protein-like and
DNA-like properties~\cite{rna}.  Due to the nature of the physical
interaction between the monomers of a RNA, aspects of the RNA
structure formation problem are considerably easier to treat than
protein folding. Also, the thermodynamic limit is more meaningful for
the RNA, which can contain over $50,\!000$ monomers.  In this paper,
we will describe the simplest effect of sequence bias to the formation
of RNA {\em secondary structures} (defined below).  We will focus on
the transition between a designed native structure and the RNA's
``molten phase'', a thermalized, collapsed phase which exists in an
intermediate temperature regime in between denaturation and
freezing. We will introduce a simplified {\em two-state} model to
describe the effect of sequence bias, following N.~G\=o's approach for
proteins~\cite{go83}.  We will solve the model exactly, and derive the
critical properties of the transition between the native and the
molten phase. The applicability of our two-state model to the
native-molten transition of designed heterogeneous RNA sequences is
verified by direct numerical enumeration and finite-size scaling
analysis.

\begin{figure}[ht]
\begin{center}
\epsfig{figure=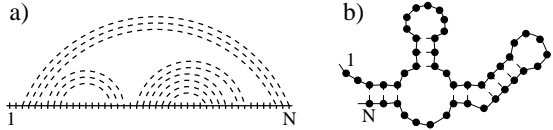,width=0.84\columnwidth}
\vspace{10pt}
\caption{Representations of the secondary structure of a RNA: (a) a
non-crossing arch diagram; (b) a helix diagram. The dashed lines
indicate base pairings.  }
\label{fig_repres}
\vspace{-10pt}
\end{center}
\end{figure}

RNA is a polynucleotide chain consisting of the four ``bases'' $A$,
$U$, $G$, and $C$. Energy of the order of several $k_B T$'s can be
gained by forming complementary pairs (i.e., $A-U$ and $C-G$) and then
stacking them in a double helical structure similar to a
double-stranded DNA.  In order to form these base pairs, the RNA will
need to bend back onto itself at various locations, resulting in a
number of helical segments.  These helices are then arranged in a
three dimensional, so-called ``tertiary'' structure, stabilized by the
much weaker interaction between the helices. Due to this crucial
separation of energy scales for the RNA, it is possible to distinguish
between the formation of ``secondary'' and tertiary structure: a RNA
secondary structure is a collection of base pairings, with the
restriction that any two base pairs $(i_1,j_1)$ and $(i_2,j_2)$ have
to be either ``nested'' (i.e., $i_1\!<\!i_2\!<\!j_2\!<\!j_1$) or
``independent'' (i.e.,
$i_1\!<\!j_1\!<\!i_2\!<\!j_2$)~\cite{wate78,zuke84,mcca90}.
Interactions violating these rules lead to structural elements which
typically cannot form simple double-helices. Thus, they are
energetically or kinetically suppressed and deemed part of the
tertiary structure.  Each such secondary structure can be represented
by a non-crossing arch diagram [see Fig.~\ref{fig_repres}(a)], where a
pairing between the bases $i$ and $j$ is indicated by a dashed line
connecting $i$ and $j$ on a stretched backbone.
Fig.~\ref{fig_repres}(b) shows an alternative representation of the
same structure; here the backbone is bent and the dashed lines are
short, in order to convey a sense of the backbone topology.  The
regions with consecutive base pairings form the above mentioned
double-helices.

To study the thermodynamic ensemble of all possible secondary
structures of a given RNA molecule, the energy of each structure needs
to be specified. Here we shall take the simplest energy function, with
$v(b,b')$ for each pairing of the bases $(b,b')$, and $v_0\!=\!-s_0T$
for each unpaired base mimicking the entropy gained from
unbinding~\cite{loop}.  Accurate energy parameters including the
effects of stacking, loops, etc.~\cite{mcca90} should be used for
predicting actual secondary structures of real RNA
molecules. Due to their irrelevance~\cite{bund99a} for the
asymptotic properties studied here we neglect them for
simplicity.  In fact, our parameters $v(b,b')$ should be viewed as
{\em coarse-grained} quantities describing the energy of pairing two
short {\em segments} of bases.

The class of RNA secondary structures is clearly hierarchical and
belongs to the class of Hartree diagrams widely used in the
self-consistent treatment of many-body quantum systems.  [In this
sense, one may regard the RNA secondary structure problem as the
Hartree theory of the ``full'' (protein-like) heteropolymer problem.]
The recursive nature of the diagrams allows efficient computation of
the exact partition function of an arbitrary sequence: Consider a
segment of bases from the positions $i$ to $j>i$ inclusive. The base
at $j$ can be either unbound or bound to any base
$k\in\{i,\ldots,j-1\}$.  For the simple energy function we have
adopted here, the partition function $Z_{i,j}$ for this segment of
bases then obeys
\begin{equation}
Z_{i,j}=Z_{i,j-1} +
\sum_{k=i}^{j-1} Z_{i,k-1}\cdot e^{-\varepsilon_{k,j}/T} \cdot
Z_{k+1,j-1},\label{eq_partfuncrec}
\end{equation}
where we take $\varepsilon_{i,j}\!\equiv\!v(b_i,b_j)\!-\!2v_0$.  The
partition function $Z_{1,N}$ of a strand of length $N$ can be computed
recursively using (\ref{eq_partfuncrec}) (with
$Z_{i,i}\!=\!Z_{i,i-1}\!=\!1$) in $O(N^3)$ time~\cite{mcca90}.

Before we discuss the effect of structure formation due to sequence
bias, we first give a qualitative description of the behavior of an
uncorrelated random RNA sequence~\cite{bund99b}.  The energetics of
this system is determined by the mean $\varepsilon_0(T)$ and standard
deviation $\delta\varepsilon$ of the pairing energies
$\varepsilon_{i,j}$. Denaturation occurs at a temperature where
$\varepsilon_0(T_d) \approx 0$, since for large and positive
$\varepsilon_0$, the unbound state is preferred~\cite{bund99a}. Below
the denaturation temperature $T_d$, the bases of the RNA are mostly
paired together. There, the system can take on two possible phases: At
very low temperatures ($T\!\ll\! \delta\varepsilon$), heterogeneity of
the sequence is important, forcing the polymer to adopt the optimal
base-pairings which minimize the total energy; this is the glass
phase~\cite{bund99b,higg96}.  At intermediate temperatures
($T_d\!>\!T\!\gtrsim\! \delta\varepsilon)$, differences in the binding
energy are less important while an average attraction between the
monomers still exists. There, the entropy of forming different
pairings becomes dominant, resulting in the molten phase.  In a
separate study~\cite{bund99b}, we will demonstrate the perturbative
irrelevance of weak sequence inhomogeneity in the intermediate
temperature regime, thereby establishing the stability and
self-consistency of the molten phase.  Furthermore, as sequence
heterogeneity is irrelevant in the molten phase, statistical
properties in this phase can be obtained from (\ref{eq_partfuncrec})
by simply taking $\varepsilon_{i,j}=\varepsilon_0$, where
$\varepsilon_0<0$ can be interpreted as an effective mean attraction.
Here, we will take the existence of such a molten phase as a
conjecture, and examine the effect of sequence bias.

To do so, we need to construct a sequence with a dominant native
structure. For simplicity, we will take the native structure to be a
hairpin with a long stem, a structure which has been studied
experimentally for oligopeptides~\cite{muno97} and short
RNA~\cite{zuo90}, although we will consider here the limit where the
stem is long.  In this native structure, the bases
$(1,2L),(2,2L-1),\ldots,(L,L+1)$ of a length $N=2L$ sequence are
paired. We call these the ``native pairs'' or ``native contacts''.
Bias towards this structure can be ``designed'' into the sequence by
choosing a {\em random} sequence for the bases $1$ to $L$ of the
molecule and then taking the second half of the molecule ($L+1$ to
$2L$) to be the exact reverse complement of the first half.  The
perfectly complementary native pairs then make the native structure
the ``ground state'' of the system.  Upon increasing temperature, the
entropy of forming non-native pairings will compete with and hence
weaken the effective bias of the native structure.  Alternatively,
this bias can be weakened by random ``mutations'' of the designed
sequence. For sufficiently weak effective bias, the RNA can ``melt''
from its native structure into any of the denatured, molten, or glass
phase, depending on the temperature and the strength of the bias.

To study the native-molten transition of the designed sequence
analytically, we shall describe the pairing energies and the bias by a
simple two-state model,
\begin{equation}\label{eq_Go-energy}
\varepsilon_{i,j} = \widetilde\varepsilon\, \delta_{i+j,2L+1} +
\varepsilon_0,
\end{equation}
for designed sequences of length $2L$.  The first term in
(\ref{eq_Go-energy}) (with $\widetilde\varepsilon <0$) mimics the {\em
additional} attraction of native pairs due to sequence design;
$|\widetilde\varepsilon|$ characterizes the ``strength'' of design
which can be controlled by the ``mutation'' process mentioned above.
The second term describes the average attraction of the ``background''
characteristic for the molten phase.  The two-state model
(\ref{eq_Go-energy}) is conceptually similar to the one introduced by
N.~G\=o in the context of protein folding~\cite{go83}.  While G\=o
proposed this model to simplify the numerical simulation of lattice
protein models, we will show that this model actually gives a
quantitative description of the native-molten transition of the RNA
secondary structure problem (\ref{eq_partfuncrec}).

Our task now is to study the system defined by
Eqs.~(\ref{eq_partfuncrec}) and (\ref{eq_Go-energy}).  We begin with a
description of the molten phase, with $\widetilde\varepsilon$ in
(\ref{eq_Go-energy}) set to zero. This phase is described by a single
parameter, $q\equiv e^{-\varepsilon_0/T}$. Due to the {\em
translational invariance} of the uniform interaction, the partition
function can be written as $Z_{i,j}=Z_0(j-i+2;q)$.  In terms of the
Laplace transform $\widehat Z_0(\mu;q)=\sum_{\ell=1}^\infty
Z_0(\ell;q) e^{-\mu \ell}$, the recursion relation
(\ref{eq_partfuncrec}) takes on the simple form $\widehat
Z_0^{-1}=e^\mu-1-q\widehat Z_0$.  Inverse Laplace transforming this
solution in the limit of large~$\ell$ using the saddle point method,
one finds the asymptotic form~\cite{wate78,dege68}
\begin{equation}\label{eq_Z0}
Z_0(\ell;q) =   A(q)\,\ell^{-\theta}e^{\mu_0(q)\ell}
\end{equation}
with $\theta=3/2$ and $\mu_0(q)=\log(1+2\sqrt{q})$.  Physically, the
partition function describes the configurational entropy of forming
different secondary structures. Each such structure can be viewed as 
a configuration of an annealed (and rooted) {\em branched polymer}%
~\cite{bund99a,lube81}, as Fig.~\ref{fig_repres} suggests. 
Note that the exponent $\theta=3/2$ is an important
characteristic of the molten phase and will play a key role in what
follows.

We now include the additional energetic bias $\widetilde\varepsilon$
in (\ref{eq_Go-energy}) due to sequence design.  For a RNA sequence of
$2L$ bases, observe that each secondary structure consists of a series
of native pairings, e.g., $(i_1, 2L-i_1+1)$, $(i_2, 2L-i_2+1)$, etc.,
separated by ``bubbles'' of lengths $\ell_k=i_{k+1}-i_k$ containing
{\em only} non-native pairings of the intervening bases.  Let the
Boltzmann weight of each native pairing be $\widetilde q\equiv
e^{-\widetilde\varepsilon/T}$, and let the restricted partition
function describing all possible non-native pairings in a molten
bubble of length $\ell$ be $W(\ell;q)$.  The total partition function
$Z(L+1;\widetilde{q},q)$ for the model (\ref{eq_Go-energy}) can then
be conveniently written in Laplace space as 
\begin{equation}\label{eq_zwrel}
\widehat Z(\mu;\widetilde q,q)=
\widehat W(\mu;q) \ \sum_{n=0}^\infty 
\left(q\widetilde q \widehat W(\mu;q) \right)^n
\end{equation}
which is easily summed.
Here, $\widehat{W}(\mu;q)$ and
$\widehat{Z}(\mu;\widetilde{q},q)$ are respectively the Laplace
transform of $W(\ell;q)$ and $Z(L;\widetilde{q},q)$.  $\widehat
W(\mu,q)$ is completely specified  by Eq.~(\ref{eq_zwrel}) and the 
``boundary condition''
\begin{equation}\label{eq_cond}
Z(L+1;\widetilde q=1,q)=Z_0(2L+1;q).
\end{equation}

Before we proceed with an analytical solution of this system, let us
observe that Eq.~(\ref{eq_zwrel}) is mathematically very similar to the
equation derived by Poland and Scheraga~\cite{pola66} describing the
{\em thermal denaturation} of perfectly complementary DNA double
strands.  The latter description in turn is a refined version of the
{\em helix coil transition} proposed by Zimm~\cite{zimm60}
[Note however that while the helix-coil transition and the DNA denaturation
transition consider {\em only} the interaction of native pairs, the
RNA folding problem considered here includes interactions between bases 
far apart along the backbone of the chain.]
The mathematical similarity can be made
clearer if one assumes (as it will turn out to be the case) that the
number of native pairings in each molten bubble is a negligible
fraction of the total number of pairings, i.e., $W(\ell; q) \approx
Z_0(2\ell-1;q)$. Then, from the form of (\ref{eq_Z0}), one can think
of $W(\ell)$ as the Boltzmann weight of a ``Gaussian polymer loop'' of
length $\ell$ in $d$-dimensions, with the fictitious dimension $d$
given by $2\theta$. Thus, the molten bubbles described by $W$ are
analogous to the denaturation bubbles in the standard denaturation
problem, or to the coiled regions in the helix coil transition. They
all represent the entropically favored phase but the origin of
these entropies is quite different: The denaturation bubbles are
formed by the configuration entropy gained by the unconstraint chains,
while the molten bubbles are driven by the ``branching entropy'' of
secondary structures within the molten phase.  In any case, we expect
that the native-molten transition belongs to the same universality
class as the denaturation transition of Ref.~\cite{pola66}, with
$d=2\theta=3$. 

The partition function $Z(L;\widetilde{q},q)$ can actually be obtained
{\em exactly} in the limit of large $L$, although the details are
somewhat tedious~\cite{bund99a}. First, by using Eqs.~(\ref{eq_zwrel})
and (\ref{eq_cond}), an {\em exact\/} expression for
$\widehat{Z}(\mu;\widetilde q,q)$ for {\em arbitrary} $\widetilde q$
can be derived.  It is straightforward to extract the reduced free
energy $f(\widetilde q,q)\equiv-(\log Z)/L$ from its singularities.
It is $-2\mu_0(q)$ with $\mu_0(q)$ as given in (\ref{eq_Z0}) for
$\widetilde q\le\widetilde q_{\mathrm{c}}$ and some other implicitly
known function $-\mu_1(\widetilde q,q)$ for $\widetilde q\ge\widetilde
q_{\mathrm{c}}$.  The critical point is at ${\widetilde
q}_{\mathrm{c}}=(3\sqrt{1+2\sqrt{q}}-1)/(\sqrt{1+2\sqrt{q}}-1)>1$.
Expanding around ${\widetilde q}_{\mathrm{c}}$ yields
$f(\widetilde{q}) - f(\widetilde{q}_c) \sim(\widetilde q-{\widetilde
q}_{\mathrm{c}})^2$, which implies a {\em continuous} phase transition
with a {\em finite jump} in the specific heat at the critical point;
thus, the specific heat exponent is $\alpha=0$. {From} the free
energy, we can easily compute the average {\em fraction of native
contacts}, $Q=-{\mathrm{d}} f/{\mathrm{d}}\ln\widetilde{q}$, which
constitutes the {\em order parameter} of the phase transition.  In the
thermodynamic limit, $Q=0$ for $\widetilde
q\le{\widetilde{q}}_{\mathrm{c}}$ and $Q=1$ for $\widetilde q \gg
{\widetilde{q}}_{\mathrm{c}}$.  Close to the critical point, $Q \sim
(\widetilde{q}-{\widetilde{q}}_{\mathrm{c}})$.

For strands of finite length $L$, this length always enters the saddle
point equation involved in the inverse Laplace transform of
$\widehat{Z}(\mu)$ in the combination $L(\widetilde
q-{\widetilde q}_{\mathrm{c}})^\nu$ with $\nu=2$~\cite{bund99a}. 
The finite-size result 
can be cast into the form $Q(L) = L^{-1/2} \,
g\left[(\widetilde{q}-{\widetilde{q}}_{\mathrm{c}})L^{1/2}\right]$ in
the vicinity of the critical point. The scaling function $g[y]$ can be
computed exactly numerically, with $g[y] \sim y$ for $y\gg 1$ 
and $g[y]\sim 1/y$
for $ y \ll -1$, with $g[y]\sim O(1)$ for $|y|\ll 1$.

In order to verify whether the above critical behavior, as derived
{from} the simplified system ~(\ref{eq_partfuncrec}) and
(\ref{eq_Go-energy}) describes the native-molten transition of
designed {\em heterogeneous} sequences, we numerically iterated
Eq.~(\ref{eq_partfuncrec}) for perfectly designed sequences%
~\cite{noglass}.  The energies $\varepsilon_{i,j}$'s were
chosen to be $-1$ for complementary pairs and $0$ otherwise.
\begin{figure}[ht]
\begin{center}
\epsfig{figure=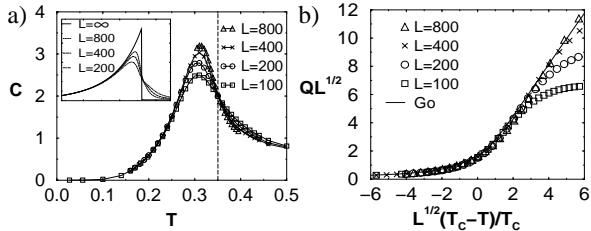,width=0.9\columnwidth}
\vspace{10pt}
\caption{Numerical results on RNA sequences with bias for a hairpin.
(a) Specific heat as a function of temperature for different system
sizes. The vertical line indicates the critical temperature; the inset
sketches the same quantity from the exact solution of the G\=o-like
model.  (b) Scaling plot of the fraction $Q$ of native contacts
with the exact solution for the scaling function \protect$g[y]$ of the
G\=o-like model as the solid line.}
\label{fig_numer}
\vspace{-10pt}
\end{center}
\end{figure}
Fig.~\ref{fig_numer}(a) shows the specific heat as a function of the
temperature for perfectly designed sequences of $200$ to $1600$ bases
averaged over an ensemble of $100$ realizations of randomness.
Direct extraction of critical exponents from this data is difficult due
to the strong correction-to-scaling effects of the expected
discontinuity ($\alpha=0$).  However, for $\alpha=0$, a good numerical
estimate of the critical temperature $T_c$ can be obtained {from} the
common intersection point of the curves at different lengths.  This is
more clearly seen from the result of the G\=o-like model
(\ref{eq_Go-energy}); see inset of Fig.~\ref{fig_numer}(a). The
fraction $Q$ of native contacts can then be used for a detailed
scaling analysis.  As shown in Fig.~\ref{fig_numer}(b), the scaling
plot of $QL^{1/2}$ versus $L^{1/2}(T_{\mathrm{c}}-T)/T_{\mathrm{c}}$
is consistent with the predicted critical behavior around the phase
transition and is well described by the scaling function $g[y]$ of the
G\=o-like model.

To summarize, we analyzed the heteropolymer structure formation
problem in the context of RNA secondary structures. The native-molten
structural transition results from a competition between the energetic
gain of native contacts and the ``branching entropy'' of the molten
phase. Critical properties can be obtained exactly after introducing
an approximate two-state model \`a la G\=o; the validity of the
approximation is verified by direct numerical calculation of designed
sequences. Throughout this study, we have neglected the effect of the
excluded-volume interaction~\cite{chen95}.  This
effect changes the value of the exponent $\theta$~\cite{pari81}, hence
changing the universality class or even the order of the phase
transition in $3$ dimensions as will be discussed
elsewhere~\cite{bund99a}; it however does not change the qualitative
physics of the competing interactions discussed here. 
The latter should be accessible experimentally using the recently developed
molecular beacon technique~\cite{bonn98}.
Finally, our study should also be relevant to the statistics of
heteroduplex formation which occurs in the hybridization of partially
complementary heterogeneous DNA strands~\cite{wetm91}. In this regard,
the structural transition discussed here resembles the physics of
similarity detection discussed previously in the context of DNA
sequence alignment~\cite{hwa96}.

We are grateful to D.~Cule who was involved in the early stages of
this study, and to discussions with M. Zuker, J.N.~Onuchic and
J.D.~Moroz. RB is supported by a Hoch\-schul\-son\-der\-pro\-gramm III
fellowship of the DAAD and TH by a Beckman Young Investigator Award.

\end{multicols}

\begin{references}
\bibitem{noglass}
The sequence is chosen to have the strongest bias so that it is far
away from the native-glass transition.
\end{references}
\end{document}